# First Principle-based Analysis of Single-Walled Carbon Nanotube and Silicon Nanowire Junctionless Transistors

Lida Ansari, Baruch Feldman, Giorgos Fagas, Carlos Martinez Lacambra, Michael G. Haverty, Kelin J. Kuhn, Sadasivan Shankar, and James C. Greer

*Abstract*—Junctionless transistors made of silicon have previously been demonstrated experimentally and by simulations. Junctionless devices do not require fabricating an abrupt source-drain junction and thus can be easier to implement in aggressive geometries. In this paper, we explore a similar architecture for aggressively scaled devices with the channel consisting of doped carbon nanotubes (CNTs). Gate all around (GAA) field effect transistor (FET) structures are investigated for *n*- and *p*-type doping. Current-voltage characteristics and sub-threshold characteristics for a CNT-based junctionless FET is compared with a junctionless silicon nanowire (SiNW) FET with comparable dimensions. Despite the higher on-current of the CNT channels, the device characteristics are poorer compared to the silicon devices due to the smaller CNT band gap.

## I. Introduction

Over the past several decades, the semiconductor industry has successfully followed Moore's law by doubling the number of transistors on a chip every two years. The International Technology Roadmap for Semiconductors (ITRS) predicts that metal-oxide-semiconductor field effect transistors (MOSFET) will reach sub-10 nm dimensions by 2018. Due to their high mobility, carbon nanotubes are attractive candidate as a channel material in aggressively scaled FETs. Carbon nanotubes were first synthesized in 1991 [1] and can be described as a sheet of a graphite monolayer rolled into a cylinder. Theoretical calculations have predicted and experiments have confirmed that single walled carbon nanotubes (SWCNTs) can exhibit either metallic or semiconducting behavior depending on the nanotube diameter and helicity of the carbon atoms along the cylinder axis [2]. Semiconducting nanotubes with small diameters are suitable for electronic devices with band gaps on the order of an electron volt [3]. Due to their one-dimensional nature and high structural quality, SWCNTs can be used as channels in high performance FETs [4,5]. Additionally, their small diameter enables excellent electrostatic control by gate electrodes [6]. The first CNT-based FET was fabricated successfully in 1998 [7]. Similar to silicon transistors, various strategies have been employed to scale down CNT-based FETs and maximize device performances, such as using a top gate [8], high-κ dielectric material [9] and a short channel length [10].

Modern transistors have reached length scales which require ultra-sharp doping concentration gradients to form junctions, thereby posing significant challenges to fabrication [11]. At nanometer length scales, formation of *p-n* junctions as required in conventional transistor designs is difficult to achieve, and random effects due to the positions of dopant atoms suggest that such *p-n* junctions cannot be routinely realized on the scale of few nanometers. The donated electrons delocalize over a distance comparable to the gate length [12], as schematically shown in Fig. 1. In a junctionless device, source, drain and channel are uniformly doped. Therefore, a junctionless design could be advantageous at the nanometer length scale [12]. Gate all around (GAA) configurations allows for better electrostatics comparing to planar counterparts [13] and there is an advantage of the cylindrical shape of CNT nanotubes which maximizes capacitive coupling between the gate electrode and the nanotube channel.

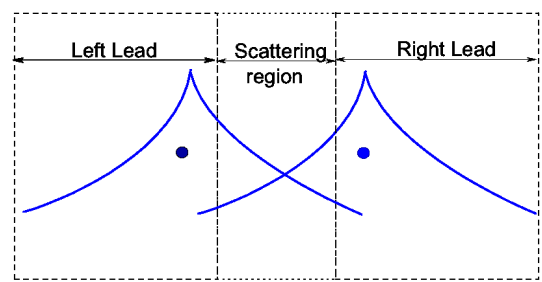

Fig. 1. Schematic description of the overlapping of the left and right leads dopants wavefunctions in the scattering region (channel). For a silicon nanowire grown along the <110> orientation with 1.15 nm diameter, the localization radius of dopants is around 1.5 nm.

Manuscript received March 12, 2013. This work was supported by Science Foundation Ireland under grant 06/IN.1/I857 and by Intel Corp. Computing is done using the Irish Center for High End Computing (ICHEC).

L. Ansari, B. Feldman, G. Fagas, and J. C. Greer are with Tyndall national institute, University College Cork, Cork, Ireland. (Lida.Ansari@tyndall.ie; BaruchF@alum.mit.edu; Georgios.Fagas@tyndall.ie; Jim.Greer@tyndall.ie).
C. M. Lacambra is with Intel Ireland, Leixlip, Ireland. (Carlos.Martinez.Lacambra@intel.com). M. G. Haverty and S. Shankar are with Intel Corporation, Santa Clara, USA. (Michael.G.Haverty@intel.com; Sadasivan.Shankar@intel.com). Kelin. J. Kuhn is with Intel Corporation, Hillsboro, USA (kelin.ptd.kuhn@intel.com).

Recently, junctionless Si nanowire (SiNW) FETs have been fabricated. Both experimental and simulation results reveal that as is common with all gate-all-around architectures, excellent electrostatic control of the gate over the channel region [11]. The ideal channel material for nanoscale junctionless transistors is at present unknown. Silicon remains a highly attractive candidate from the standpoint of both physical and economic arguments. In this paper, we present atomic scale simulation of junctionless CNT-FETs and compare their performance to SiNW junctionless transistors of comparable dimensions.

## II. METHODOLOGY

Material properties such as mobility and effective masses are substantially different at the few nanometer length scale compared to the bulk, and their determination requires an explicit treatment of the electronic structure in nanoscale devices. Also, a description of charge transport needs to be based on quantum mechanics as transistor dimensions are scaled down to dimensions comparable to the Fermi wavelength.

A semiconducting (10,0) SWCNT is used to construct a junctionless CNT-FET with *p*-type and *n*-type channels formed by substitutional doping with boron and nitrogen, respectively. A doping concentration of 1.1% is used for both *n*- and *p*-type. It has been reported that the boron concentration of individual nanotubes is of the order of 1-5% [14], hence, our SWCNT channels are at the lower end of this range but still are highly doped compared to conventional transistors. Note that this high doping level is consistent with the junctionless transistor operation [11,12]. With nitrogen dopants, doping concentrations up to 8.4% have been achieved previously in experiments [15].

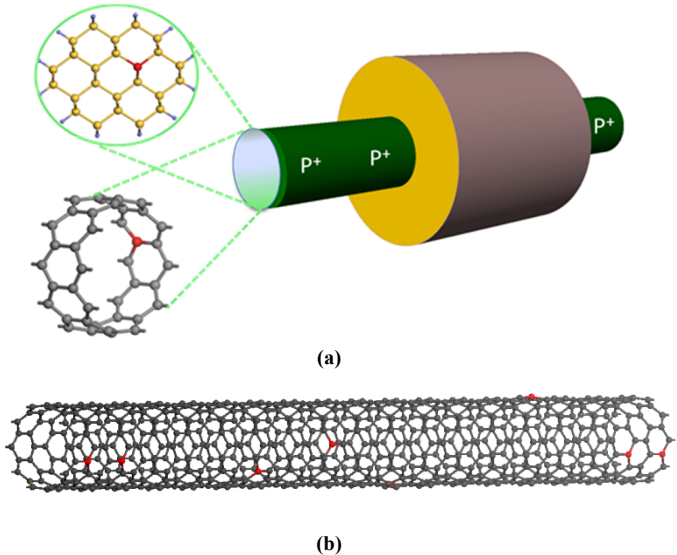

Fig. 2. (a) Schematic structure of junctionless transistor with gate all around configuration. Also shown: the cross-section of SiNW along <110> and semiconducting (10,0) SWCNT with a substitutional dopant which are used as channel material in this study. (b) Doping profile of the randomly doped (red spheres) *p*-type CNT in source, drain and channel.

We consider a structure with gate length ($L_g$) of 3.2 nm, tube length of 7.6 nm, and tube diameter of 0.79 nm, consisting of a total of 720 atoms. Fig. 2(a) shows the schematic structure of SiNW or SWCNT based junctionless transistor. As it is shown in figure 2(b) the dopants are randomly distributed in source, channel and drain. The doping profile of the nitrogen doped CNT was considered to be the same as boron doped CNT.

Boron and phosphorous are used to make a *p*-type and *n*-type junctionless SiNW channels, respectively. The doping concentration for the SiNW channels is $1.5 \times 10^{20}$ cm$^{-3}$ (eight dopant atoms are distributed within the 800-atom supercell). Equilibrium dopant concentrations of $1.5 \times 10^{20}$ cm$^{-3}$ have been recently measured for SiNWs [16]. The length of the SiNW is 7.7nm with $L_g$ of 3.1 nm. The cross-section of a SiNW along the <110> crystallographic direction is shown in Fig. 2(b).

To determine the electronic structure, we used density functional theory (DFT) in a numerical atomic orbital basis, as implemented in the OpenMX program [17]. DFT is applied with the local density approximation (LDA) and with norm conserving pseudopotentials. The basis set for each type of atom is chosen as the optimized orbitals from a quadruple-zeta orbital set [18,19]. The structural configurations for the doped CNTs and SiNWs are obtained by relaxing the simulation cell to minimize the total energy and relaxing the atomic structure until the maximum force component is less than 0.01 eV/Å.

In order to investigate the effects of a single dopant atom on the CNT, we studied the net charge difference between an intrinsic and *p*-type CNT. Fig. 3(a) illustrates the difference between the Mulliken charges for an intrinsic CNT and boron doped CNT, as given below:

$$\Delta Q_i^M = Q_i^M - Q_i^{M,0}, \qquad (1)$$

where *i* is the atom index, $Q_i^M$ is the Mulliken charge of atom *i* in the boron doped CNT, and $Q_i^{M,0}$ is the Mulliken charge in the intrinsic tube. Mulliken population analysis is a post-processing step for localized-basis calculations yielding a charge associated with each atom [20]. The net charges $\Delta Q_i^M$ are indicated by red spheres and the net charge on a dopant site is highlighted by a blue circle. As can be seen in this figure, the charge is transferred to the three neighboring atoms of the dopant labeled as 1, 2, and 3. The Mulliken charges around the dopant atoms (boron or nitrogen) indicate the presence of an acceptor or a donor complex as indicated in Fig. 3(b).

We implement the GAA architecture by introducing an external gate bias in the Kohn-Sham self-consistent equations [21]. The electronic structure is calculated for various values of gate voltages using a version of the OpenMX program we modified for this purpose. Fig. 4 shows an isopotential surface of the GAA configuration for the SiNW junctionless devices. The oxide surrounding the CNT and SiNW is modeled as a continuum described by the dielectric constant for hafnium oxide, $\varepsilon_{HFO_2} = 25$.

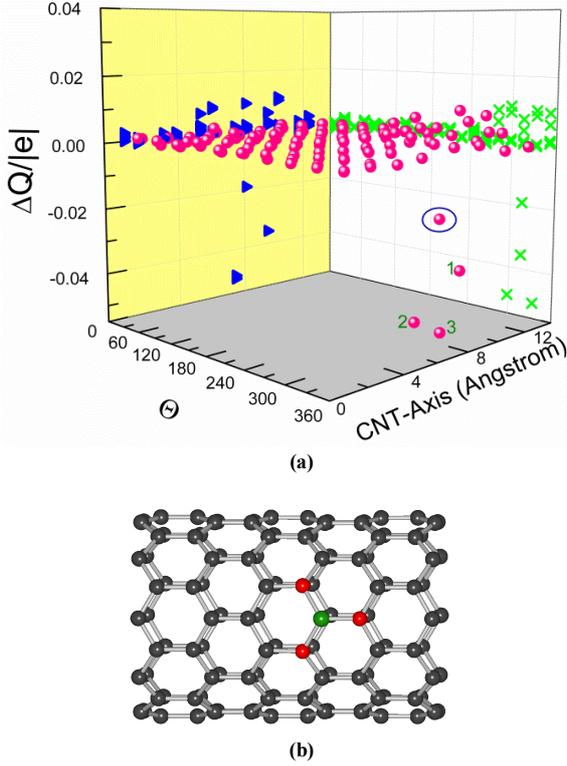

(a)

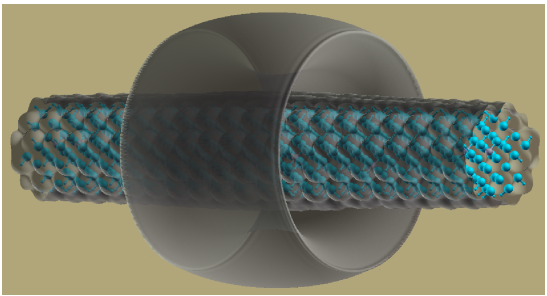

(b)

Fig. 3. (a) Mulliken charge difference (red spheres) between boron doped and intrinsic CNT along the tube axis and the angular position on the same periphery (θ). Blue triangles display the sum of the radial charge as a function of the CNT axis and green crosses indicate the angular distribution of the charges. (b) Neighboring atoms of one dopant (green sphere) presenting a doping cluster (together with red spheres).

The band structure of a CNT supercell consisting of 120 carbon atoms doped with one substitutional boron atom (0.83% doping concentration) is shown in Fig. 5(a). As can be seen, the Fermi level ($E_F$) is *inside* the valence band. Considering the same doping concentration for nitrogen doped CNT, the Fermi level is *inside* the conduction band. Hence the behavior is appropriate for *n*-type (or correspondingly *p*-type for boron doping). The difference between $E_F$ and the closest band edge for 1.1% boron doped SWCNT is 156 meV and for 1.1% nitrogen doped SWCNT is 147 meV consistent with the high doping levels introduced.

The band structure of boron doped SiNW with one dopant in every second unit cell is shown in Fig. 5(b) for comparison; note the larger band gap relative to bulk silicon resulting from quantum confinement in the nanowire. Note that approximate exchange correlation functions typically used within DFT method tends to underestimate bandgaps.

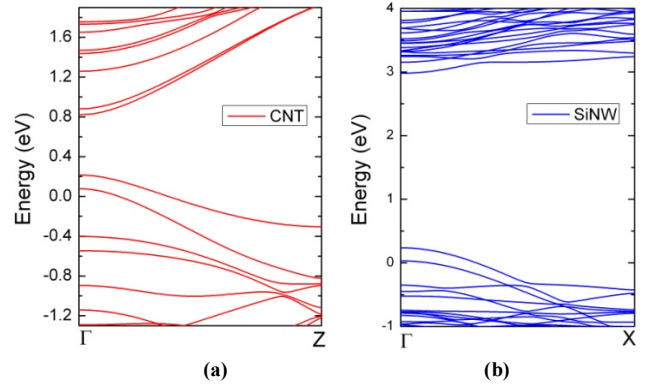

| (a) | (b) |

Fig. 5. Band structure of (a) carbon nanotube, and (b) silicon nanowire, both doped with boron. Zero energy is taken to be the Fermi energy.

The source, drain, and channel region are described by the DFT/LDA Hamiltonian matrix represented in the localized atomic orbital basis. The Hamiltonian constructed in terms of a localized basis permits a division of the system into left lead, right lead and scattering region (channel). There is no direct coupling between the leads and the leads are taken to be periodic structures.

The transport code TiMeS [22] is used to solve the scattering equations for injected electron and to determine the carrier transmission, $T(E,V)$, as a function of incident electron energy and gate potential. In this formalism, several possible Bloch waves with the same energy can propagate through the leads [23,24] and the matrix Green's function method is employed to tackle the multi-channel scattering problem. Upon obtaining the transmission, the drain-source current is calculated using the Landauer formula as follows [25]:

$$I_{DS} = \int i(E)dE,$$
$$i(E) = \frac{e}{h}T(E,V)[f_L(E) - f_R(E)], \quad (2)$$

where $f_L(E)$ and $f_R(E)$ are Fermi distribution functions in the left lead and right lead, respectively:

$$f_L(E) = \left[1 + exp\left(\frac{E - E_f}{kT}\right)\right]^{-1},$$
$$f_R(E) = \left[1 + exp\left(\frac{E - E_f - qV_{DS}}{kT}\right)\right]^{-1}. \quad (3)$$

Note that the well known deficiency of DFT/LDA theory in describing heterostructure interface alignments leading to typically overestimation of electron currents in transport formalisms is not relevant for these simulations, as there is no material interface between the leads and the channel region. Another reason for the higher current could be due to absence of phonon effects in the model currently.

Fig. 4. Isopotential surface for gated SiNW.

We have refined our electronic transport code, TIMES, to perform self-consistent non-equilibrium Greens function (NEGF) calculations in a modular fashion. Following the method of Ke et al [26] allows us to implement the NEGF independently of the electronic structure step. The channel charge density matrix is calculated based on the DFT Hamiltonian with self-energies representing semi-infinite leads in our electronic transport code. This density is used in the DFT to extract the new device region Hamiltonian matrix. This process iterates till the charge density converges. In summary, the NEGF self-consistent approach enabled us to perform the calculation for a realistic open system and the self-energies allows inflow and outflow of electrons on the explicit device region to equilibrate with the leads.

### III. RESULTS AND DISCUSSIONS

Figs. 6(a) and 6(b) show the calculated $I$-$V_{DS}$ characteristics for the $n$- and $p$-type CNT junctionless transistors, respectively, without application of the NEGF self-consistent approach. These devices are "ON" for $V_{GS} = 0$ V (no explicit metal gate is introduced, hence no metal work function offset is included in the gating voltage), and they are turned "OFF", based on the pinch-off principle when gate potential causes a sufficiently large barrier in the gate region.

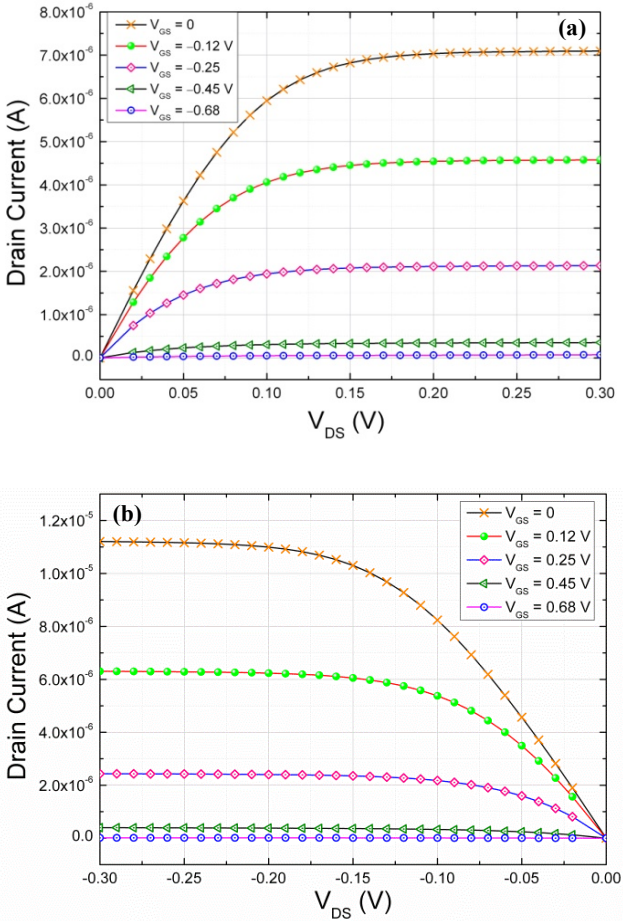

Fig. 6. Current-voltage characteristic for (a) $n$-type and (b) $p$-type CNT junctionless transistors.

The current saturation mechanism in the conventional FETs is defined using drift velocity saturation or channel pinch-off. However, for junctionless devices the current saturation is determined by band alignment and the maximum carrier flux from the source. The current saturation mechanism for a p-channel junction-less FET at zero gate voltage is shown in Fig. 7. Due to the high doping concentration, the Fermi level is inside the valence band. As seen in the figure, the carrier flux from right lead to left lead increases as the drain voltage, $V_{DS}$, increases. However, upon reaching saturation, $V_{DS\text{-}sat}$, any further increase in the drain-source voltage does not yield a further increase in the carrier flux as there is no additional density of states available for additional charge carriers to be injected from the leads.

Using the following analytical formula we calculated the "ON" current values at zero gate voltage:

$$I_{ON} = \frac{2e^2}{h} V_{DS} N_{CH}, \qquad (4)$$

where $N_{CH}$ is the number of channels. This *simple estimation* (following from Eq. 2 in the low-temperature, low bias ballistic transport regime [25]) yields an "ON" current equal to $3 \times 10^{-5}$ A for the boron-doped CNT at $V_{DS} = 0.2$ V. The simulated "ON" current value is $1.16 \times 10^{-5}$ A. Considering the fact that the number of open channels at Fermi level is equal to 2, the calculated transmission is less than 2 due to any scattering that is present in the device. Also there is a smearing around the Fermi level considering the $T = 300$ K Fermi distribution in the lead regions, with these considerations taken, there is reasonable agreement between the simple model and our simulation results.

Similarly for the "OFF" current, Eq. 2 using the Wentzel-Kramers-Brillouin (WKB) method for tunneling through a square barrier leads to the following approximation:

$$I_{OFF} = G_{OFF} V_{DS}, \qquad (5)$$

where

$$G_{OFF} = \frac{2e^2}{h} T_t N_{CH} \qquad (6)$$

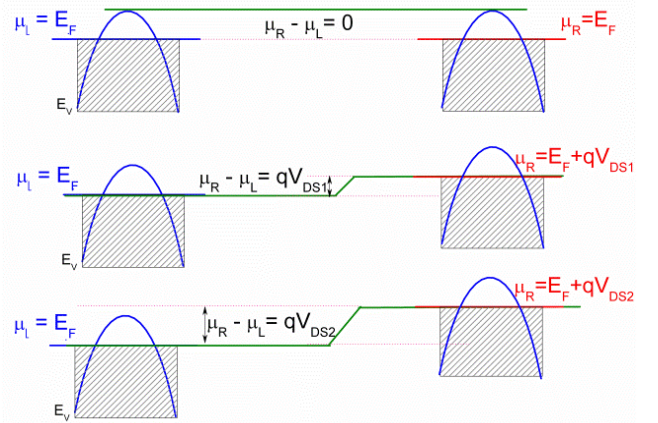

Fig. 7. Current saturation mechanism for $p$-channel devices ($V_{DS2} > V_{DS1}$).

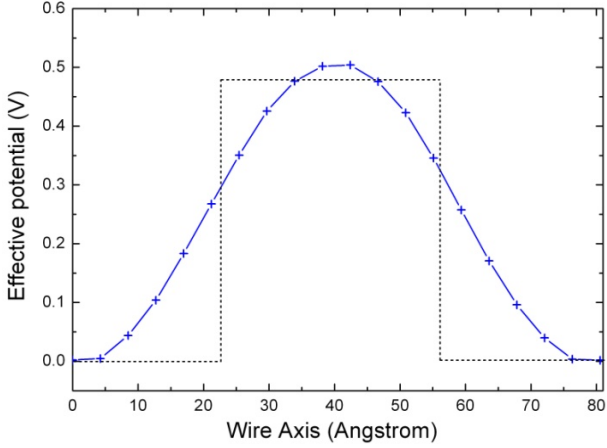

Fig. 8. Effective potential according to average Mulliken charges along the wire transport direction at $V_{GS} = 0.64$ V. Dashed line shows a square potential barrier.

and

$$T_t = \frac{exp\left(-2\sqrt{\frac{2m^*}{\hbar^2}eV_{GS}L_g}\right)}{\left(1+\frac{1}{4}exp\left(-2\sqrt{\frac{2m^*}{\hbar^2}eV_{GS}L_g}\right)\right)^2}. \quad (7)$$

The calculated hole effective mass for (10,0) SWCNT is $0.083m_0$ where $m_0$ is free electron mass. An effective potential determined from the averaged Mulliken charges without NEGF self-consistent approach is extracted and plotted in Fig. 8. If we consider a square potential barrier approximation (dashed line in Fig. 8) and use the WKB formula for tunneling the "OFF" current for a 3.2 nm gate length and 0.44 V gate voltage is predicted to be $2.36\times10^{-8}$ A, which is in good agreement with the simulated results of $I_{OFF} = 1.65\times10^{-8}$ A. However, using the self-consistent approach the "OFF" current reduces to $4.52\times10^{-10}$ A.

Figs. 9(a) and 9(b) show the difference in $I$-$V_{GS}$ characteristic obtained using NEGF self-consistent and non-self-consistent approaches for the $p$-channel SiNW and CNT devices, respectively. Using the NEGF transport algorithm and DFT, the transmission is found to be suppressed around the Fermi level compared to non-self consistent results. Hence, there is a corresponding reduction in the current as can be seen in Figs. 9(a) and 9(b). This effect is due to the fact that the charge distribution in the channel is depleted toward the leads when solved for self-consistently thereby yielding a higher barrier. For example, the total value of the net charges depleted from the channel region of a $p$-type SiNW toward the leads is 0.71|e| for self-consistent NEGF at $V_{GS} = 0.94$ V.

The $I$-$V_{GS}$ characteristics in a wider bias range are compared in Figs. 10(a) and 10(b), respectively, for $p$-channel and $n$-channel SiNW and CNT. The SiNW is doped with boron (phosphorous) to make a $p$-type ($n$-type) channel. As can be seen in these figures, the "ON" current for CNT is higher than for the SiNW. The reason is due to the larger number of open channels for CNT compared to SiNW resulting in a higher transmission at the Fermi level. Simulation results show slightly higher scattering if we consider larger disordered contacts which suppress both $I_{ON}$ and $I_{OFF}$ by similar amounts. More importantly, as higher gate voltages are applied to CNT, due to its smaller band gap and band gap narrowing caused by the formation of dopant band, valence-conduction band coupling appears in the applied bias window and as a result the "OFF" current increases again at higher bias voltages, when the top of the valence band in the channel aligns in energy with the bottom of the conduction band in the source. Similar behavior has been reported in conventional CNT-based FETs experimentally [25,26]. Much smaller "OFF" current can be achieved for SiNWs, since there is a larger band gap for SiNW due to the quantum confinement effect and the valence-conduction band coupling does not appear for higher bias points.

The effect of valence-conduction coupling in CNT channels affects the subthreshold slope of the devices. The latter is defined as the inverse of the slope of the log of drain-source current versus gate voltage at voltages below threshold:

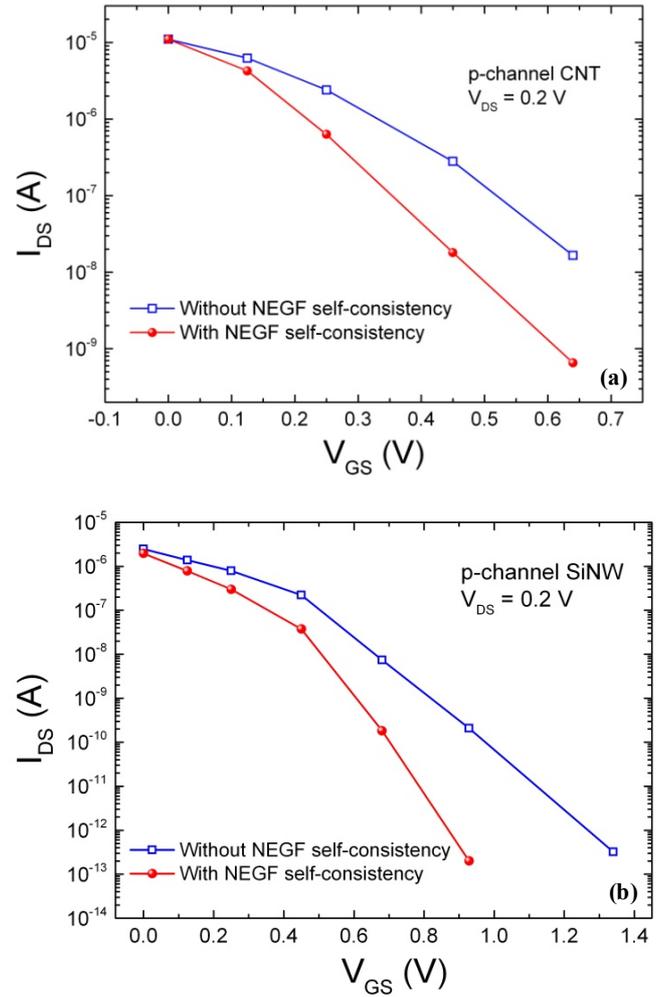

Fig. 9. $I$−$V_{GS}$ characteristic and comparison between NEGF self-consistency and non NEGF for (a) $p$-channel CNT junctionless transistor, and (b) $p$-channel SiNW junctionless transistor.

$$SS = \left(\frac{d(\log_{10}(I_D))}{dV_g}\right)^{-1}. \quad (8)$$

Based on $I-V_{GS}$ characteristics, the subthreshold slope for the *p*-channel CNT junctionless FET is 134 mV/dec and for the *n*-channel is 130 mV/dec using the self consistent approach while these values for *p*-channel and *n*-channel SiNW devices are 84 mV/dec and 89 mV/dec, respectively. The ideal subthreshold slope for a long-channel MOSFET at room temperature is 60mV/dec, whereas, in trigate inversion mode devices and short-channel junctionless transistors typical values exceed 70 mV/decade [27]. Recently, a silicon based junctionless FET with 3 nm channel length has been fabricated [28] and their experimental results demonstrated that extremely short channel device performs very well; more than $10^6$ on- off ratio and 95 mV/dec subthreshold swing for their planar structure with 3nm channel length and 1 nm film thickness.

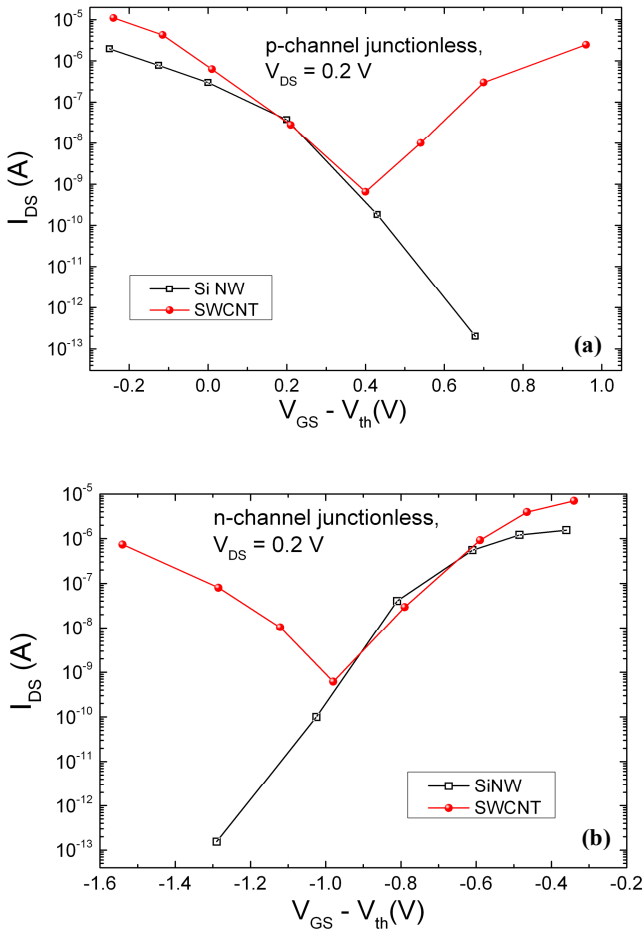

Fig. 10. Comparison between the $I-V_{GS}$ of (a) *p*-channel and (b) *n*-channel CNT and SiNW junctionless transistors. The channel length is similar in both devices.

## IV. CONCLUSIONS

We have performed first-principle transport calculations based on a self-consistent DFT + NEGF method for junctionless GAA semiconducting SWCNT FETs and SiNW FETs. Including self-consistency in the Green's function calculation of the device characteristics suppresses the off-current by allowing net charge depletion from the channel to the source and drain regions. This method allowed us to compare the performance of SiNW-based junctionless FET and CNT-based device in terms of current-voltage characteristics and subthreshold slope. According to our simulations, devices made of CNT channels are predicted to yield higher on-currents. However, due to the narrower band gap of CNTs compared to SiNWs, coupling appears between the valence band of the channel and the conduction band of the source at high bias voltages. As a result the "OFF" current increases again thereby degrading the subthreshold slope for junctionless CNT devices. On the other hand, the SiNW junctionless FET performs well even at this atomic scale limit, turning off with source-drain leakage $I_{OFF} < 10^{-6} \times I_{ON}$ for very short gate length and subthreshold slope of 84 mV/dec and 89 mV/dec for *p*-channel and *n*-channel, respectively.